\documentclass[11pt]{article}
\usepackage{coling2020}
\usepackage{times}
\usepackage{url}
\usepackage{latexsym}

\usepackage{multirow}
\usepackage{graphicx}
\usepackage{subfigure}
\usepackage{tabularx}
\usepackage{hhline}
\usepackage{array}
\usepackage{amssymb}
\usepackage{amsfonts}
\usepackage{amsmath}
\usepackage{mathrsfs}
\usepackage{color}
\usepackage{booktabs}
\usepackage{amssymb}
\usepackage[T1]{fontenc}
\newtheorem{definition}{Definition}

\makeatletter
\def\thanks#1{\protected@xdef\@thanks{\@thanks
        \protect\footnotetext{#1}}}
\makeatother

\colingfinalcopy 

\title{A General Framework for Debiasing in CTR Prediction}

\author{Wenjie Chu$^{\dagger*}$\thanks{$^\dagger$Contribute equally.} \\
  {\tt \small{wjchu@pku.edu.cn}} \\\And
  Shen Li$^\dagger$\thanks{$^*$Work performed when Wenjie worked as an intern in Alibaba.} \\
  {\tt \small{jingfu.ls@alibaba-inc.com}} \\\And
  Chao Chen$^\dagger$ \\
  {\tt \small{cc201598@alibaba-inc.com}} \\\AND
  Longfei Xu$^\dagger$ \\
  {\tt \small{longfei.xl@alibaba-inc.com}} \\\And
  Hengbin Cui \\
  {\tt \small{alexcui.chb@alibaba-inc.com}} \\\And
  Kaikui Liu \\
  {\tt \small{damon@alibaba-inc.com}} \\\AND
  Alibaba Group
}

\date{}

\begin{document}
\maketitle


\begin{abstract}
Most of the existing methods for debaising in click-through rate (CTR) prediction depend on an oversimplified assumption, i.e., the click probability is the product of observation probability and relevance probability.
However, since there is a complicated interplay between these two probabilities, these methods cannot be applied to other scenarios, e.g. query auto completion (QAC) and route recommendation.
We propose a general debiasing framework without simplifying the relationships between variables, which can handle all scenarios in CTR prediction.
Simulation experiments show that: under the simplest scenario, our method maintains a similar AUC with the state-of-the-art methods; in other scenarios, our method achieves considerable improvements compared with existing methods.
Meanwhile, in online experiments, the framework also gains significant improvements consistently.
\end{abstract}


\section{Introduction}

The prediction of Click-Through Rate (CTR) is to estimate the probability that an item will be clicked by a user in a certain scenario.
CTR is one of the most critical parts of a recommender system, especially in industry, e.g., online shops and online advertising \cite{cheng2016wide,guo2017deepfm,lian2018xdeepfm,zhu2019joint}.
For most recommender systems, the aim is to maximize revenue and user experience. 
Thus, an item with the largest predicted CTR and advertising rates will be placed in the top \cite{ling2017model,zhu2017optimized,zhou2018deep}.

\begin{figure}[t]
    \centering
    \includegraphics[width=0.6\textwidth,keepaspectratio]{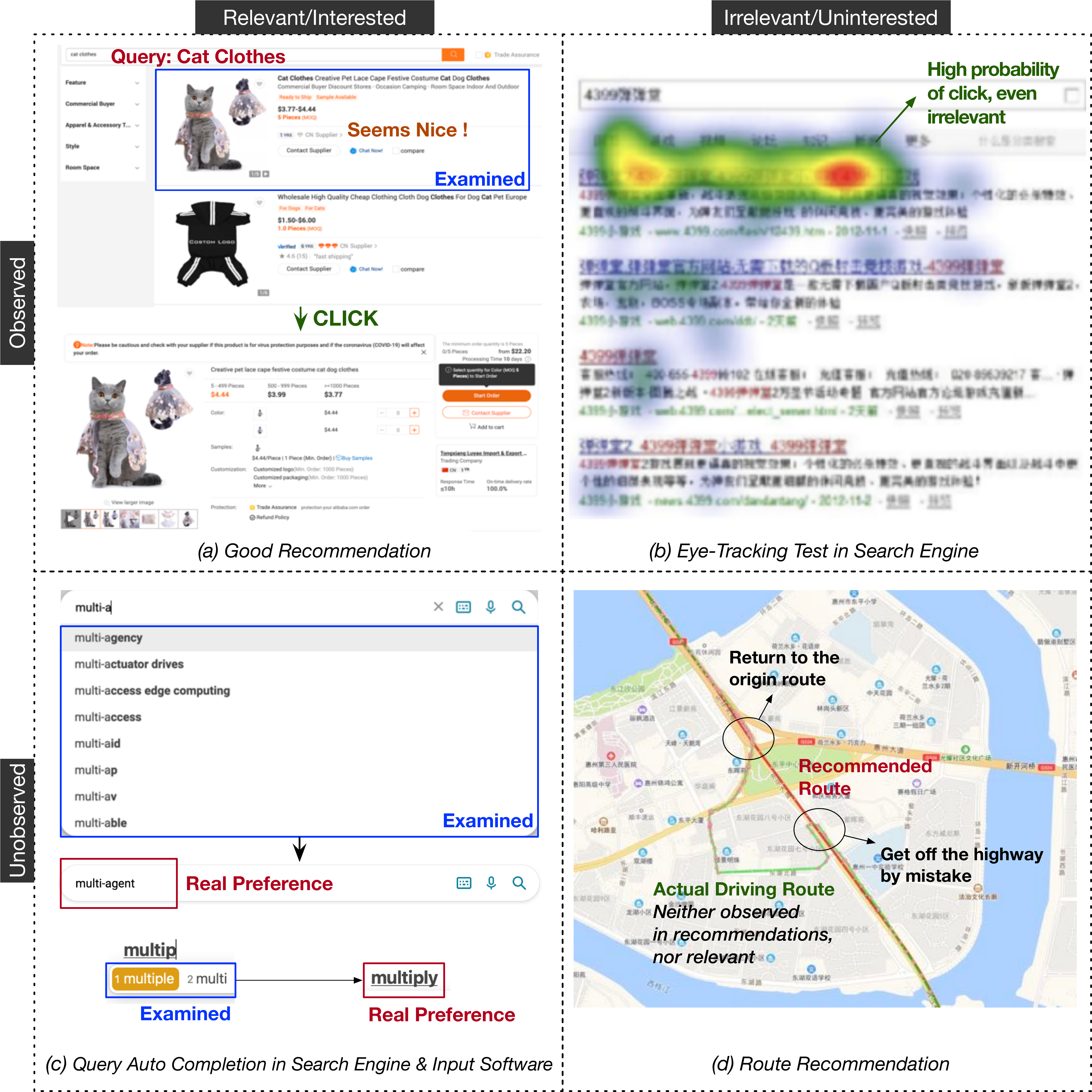}
    \caption{Four kinds of scenarios when a "CLICK" operation generates.}
    \label{fig:scen}
    \vspace{-0.5em}
\end{figure}

Two significant processes affect CTR prediction models.
In the training phase, a CTR prediction model is trained based on the information from user-item interaction, which is recorded in an online recommender system, e.g., a user viewed and clicked on an item. 
In the inference phase, the trained model is deployed to a real-time recommender system to predict the CTR of each item. 
One crucial problem of these procedures is that the user-item interaction is affected by the positions that items are displayed, i.e., an item in the top is easier to be clicked than others \cite{richardson2007predicting,guo2019pal}.
A similar phenomenon is also shown in experiments of eye-tracking, where users pay more attention to items at the top of a page than other parts (Figure \ref{fig:scen} (b)) \cite{liu2016predicting}.
Therefore, the training data collected from historical clicking suffers positional bias.
In fact, bias also includes font size, item color, the length of the title, etc.

Since positional bias is a typical bias in most situations, some approaches have been proposed to alleviate positional bias \cite{haldar2020improving,ling2017model}.
A popular method is splitting all features into a position feature and others in which two models handle two parts, respectively. 
An actual position is provided in training, and in inference, the trained model does not leverage the position feature.
With this structure, \newcite{guo2019pal} designed a PAL framework to conduct online inference without position information. \newcite{zhao2019recommending} propose a shallow side tower to learn selection bias which is learned offline and ignored online.
An assumption are widely used in most of debiasing methods where the click Bernoulli variable $y$ depends on two hidden variables \cite{richardson2007predicting}: 
\begin{equation}
    p(y = 1|x, pos) = p(s = 1|x, pos)p(y = 1|x, pos, s),
    \label{equation:origin assumption}
\end{equation}
where $p(y = 1|x, pos)$ represents the probability that a user clicks on the $pos$-th item in a specific situation $x$ including user information and context information, e.g., search query, time, location, etc. 
$p(s = 1|x, pos)$ can be considered as the probability that an item is examined. 
$p(y = 1|x, pos, s)$ is the probability that an examined item is clicked, i.e., an estimation of the relevance between the user and the item.
Furthermore, Formula \ref{equation:origin assumption} usually be transformed into
$p(y = 1|x, pos) = p(s = 1|pos)p(y = 1|x, s)$ \cite{zhao2019recommending},
where the examination probability depends only on position information $pos$.
This modification can be used to alleviate positional bias. $p(s = 1|pos)$ models bias in training phase and scores of items in online systems only depend on $p(y = 1|x, s)$ which is not affected by position.

However, since the click probability is only the product of examination probability and relevance probability, the assumption is oversimplified and insufficient to model the rich interaction between position and other information \cite{huang2021deep}.
How to estimate the probability if items are chosen but not listed in candidates?
In fact, there are four scenarios given different users and items. 
As shown in Figure \ref{fig:scen} (a), an item is clicked when a user sees it and thinks it is interesting. 
In Figure \ref{fig:scen} (b), items at the top have a higher probability of being clicked than others whether they are relevant to users or not \cite{liu2016predicting}. 
The above methods can handle these two situations, but the rest is beyond the assumption. 
After enabling QAC (Figure \ref{fig:scen} (c)), a user can select the candidates suggested by QAC or input other words via the keyboard. 
In other words, it is possible for a user to click items that are relevant but not seen by the user.
In Figure \ref{fig:scen} (d), a navigation app recommends a route along the highway, but the user gets off the highway by mistake due to watching the route carelessly, which can be regarded as choosing an unseen and irrelevant item.

To deal with all four scenarios: 1. observed \& relevant, 2. observed \& irrelevant, 3. unobserved \& relevant, and 4. unobserved \& irrelevant, we propose a general framework for debiasing in CTR prediction based on a perspective of the probabilistic graphical model.
The probability of an item being clicked by a user is determined by item information, the user's interest, the position of the item, and the bias of all contexts, no matter whether the item is observed or relevant to the user.
In the framework, two neural networks are designed to estimate the interest of the user and the transition probability of clicking, respectively.
Both of them are trained offline, and only the model predicting the interest is used online.

In simulation experiments, the framework maintains similar AUC with the state-of-the-art methods when an item is clicked if and only if it is observed and relevant to a user. The framework achieves considerable improvements compared with existing methods in other scenarios.
Furthermore, not only can positional bias be estimated by the framework, but also other biases can be modeled, such as title length.
The advantages are also be observed in online experiments.

Thus, the contribution of this paper is three-fold:
\begin{itemize}
    \item We analyze and redefine the debiasing problem in CTR prediction and split it into four scenarios.
    \item We introduce a general debiasing framework based on a probabilistic graphical model and successfully apply it to all scenarios which existing methods cannot cover fully.
    \item Offline and online experiments demonstrate that the novel framework gains a remarkable improvement compared with the state-of-the-art methods.
\end{itemize}

\section{Methodology}
\begin{figure*}[t]
    \centering
    \includegraphics[width=0.9\textwidth,keepaspectratio]{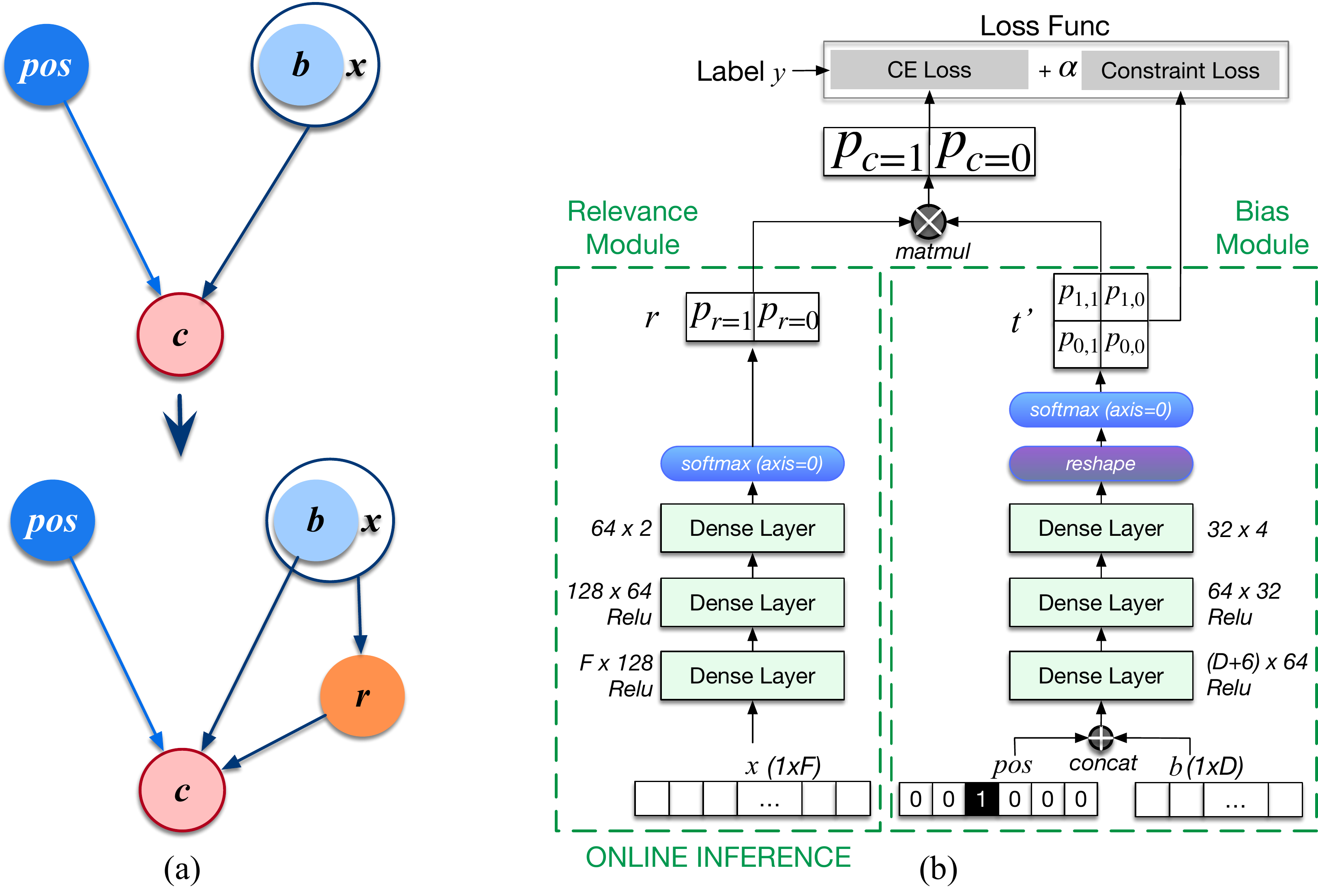}
    \caption{An overview of the implicit intention network. (a) The probability transition diagram. (b) The architecture of the proposed IIN.}
    \label{fig:module}
\end{figure*}

\subsection{Notation}
Suppose the click dataset is represented as $D=\{(x_i,pos_i,b_i,y_i)|1\leq i \leq N\}$. For sample $i$ in $D$, $x_i$ is the feature vector includes user profile, item features and context information, $pos_i$ is the position information which is usually the ranking of the item, $b_i$ is a subset of $x_i$ that contains bias information, $y_i$ is the label which is 1 if user clicks on the item and 0 otherwise. For the convenience of presentation, we use $x$, $pos$, $b$, $y$ to denote variables corresponding to feature vector, position information, bias information and label respectively.

\subsection{Implicit Intention Network}
\subsubsection{Probabilistic Graphical Model Perspective}
Implicit Intention Network (IIN) is essentially a probabilistic graph model based on the assumption that no matter whether an item is observed or not, and whether an item is relevant to the user or not, it could be clicked by the user. As shown in Figure~\ref{fig:module}.(a), the probability of an item being clicked $y$ is determined by $pos$, $b$ and the unobserved or latent variable $r$, where $r$ is the relevance between the user and the item. The probability graph model can be formulated as follows: $p(y|x,pos,b) = p(y|r,pos,b)p(r|x)$, where $p(y|r,pos,b)$  is the learnable probability transition matrix from $r$, $pos$, $b$ to $y$, and $p(r|x)$ is the real learning goal of the model which denotes the mapping function from $x$ to relevance $r$ that will be used for online sorting.

Furthermore, assuming that $r=1$ indicates the item is relevant to the user, and $0$ otherwise, then the following derivation can be made:
\begin{equation}
    \begin{aligned}
    P(y=1|x,pos,b)&=\sum_{r}P(y=1|r,pos,b)P(r|x)
    \\&=P(y=1|pos,b,r=0)P(r=0|x)
    \\&+P(y=1|pos,b,r=1)P(r=1|x)
   \end{aligned}
\end{equation}
The above formula can intuitively reflect our hypothesis. Take route recommendation as an example, where an item is a route between the origin and destination, and $y=1$ means the user drives along with the item. 
When the item is related to the user, if its sort position is visible, it has a click probability, which is easy to understand. 
If the sorting position is not visible, the user may still click on the item. 
For example, the user is familiar with the area, and even though the item is not seen, he still chooses the item to drive. 
When the item is not related to the user, there is a click probability if its sorted position is visible. 
For example, the user habitually drives along the first recommended route, but he is dissatisfied with the route after driving.
If the sorted position is not visible, there is still a click probability.
For example, the user accidentally went wrong and walked onto the item.

\subsubsection{Framework}
We can use shallow neural networks to achieve IIN. 
As shown in Figure~\ref{fig:module}.(b), the framework mainly contains 3 modules, a relevance module, a bias module, and a combined module.
Assuming that $r=1$ indicates the item is relevant to the user, and $0$ otherwise, then the following derivation can be made:
\begin{equation}\label{eq:p1}
    \begin{aligned}
    P(y=1|x,pos,b)&=\sum_{r}P(y=1|r,pos,b)P(r|x)
    \\&=P(y=1|r=0,pos,b)P(r=0|x)
    \\&+P(y=1|r=1,pos,b)P(r=1|x)
    \end{aligned}
\end{equation}
\begin{equation}\label{eq:p2}
    \begin{aligned}
    P(y=0|x,pos,b)&=P(y=0|r=0,pos,b)P(r=0|x)
    \\&+P(y=0|r=1,pos,b)P(r=1|x)
    \end{aligned}
\end{equation}

The relevance module aims to fit the function $p(r|x)$ by achieving relevance prediction with a fully connected neural network and a softmax layer. The input of this module is $x$. The output of this module is $r$, which is a two-dimensional vector and the first dimension represents $P(r=1|x)$ and the second dimension represents $P(r=0|x)$. The softmax layer guarantees $P(r=1|x)+P(r=0|x)=1$.

The bias module aims to fit the function $P(y|r,pos,b)$, which equals to four expanded functions as shown in  Eq.~\ref{eq:p1} and Eq.~\ref{eq:p2}.
Specifically, let $|b|$ and $|pos|$ represent the number of different values of $b$ and $pos$, respectively, 
then we can directly initialize a matrix of $|b|\times|pos|\times2\times2$ as the probability transfer matrix $t$. 
Given $b$, $pos$ and $r$, the predicted $y$ can be obtained from $t$. 
However, in general, $|pos|$ is large and $b$ contains complex bias information, which takes more values, even continuous. 
To avoid this, the bias module takes $pos$, $b$ as inputs and generates a $2\times2$ matrix $t'$ with a fully connected neural network and a softmax layer.
$t'$ is considered as the probability transition matrix from $r$ to $y$. This module divides $t$ into two parts: variables with more or even infinite values are modeled by a neural network, and variables with limited and few values are modeled by $t'$.
In this way, on the one hand, we can ensure the learnability of $t$; on the other hand, we can observe whether the learning of bias is reasonable through $t'$.
From another point of view, this module produces different $t'$ under different $pos$ and $b$, which is consistent with the idea that different bias information has different functions. 
The softmax layer guarantees $P(y=1|r,pos,b)+P(y=0|r,pos,b)=1$.
Consider $r$ as a matrix of $1\times2$, the combined module performs $r\times t'$ to generate predicted $y$, and calculates cross entropy loss $L_e$ for optimizing.

\subsubsection{Deterministic Constraint}
The dimension representing $P(r=1|x)$ in $r$ will be used for online ranking. Because there are uncertain factors in the training stage, such as random initialization, the relevance module can not determine which dimension of $r$ represents $P(r=1|x)$. 
To eliminate this uncertainty, a constraint on $t'$ is constructed, which is based on the assumption that for all $pos$ and $b$, $P(y=1|r=1,pos,b) > P(y=1|r=0,pos,b)$. 
Suppose that $t'_{00}$ is $P(y=1|r=1,pos,b)$ and $t'_{01}$ is $P(y=1|r=0,pos,b)$, then the constraint loss can be denoted as:
\begin{equation}
    L_c=\textbf{relu}(t'_{01}-t'_{00}),
\end{equation}
and the total loss of IIN is:
\begin{equation}
    L=L_e+\alpha L_c,
\end{equation}
where $\alpha$ can be set to a large value to ensure $L_c=0$ before the training is completed.

\section{Evaluation}
\subsection{Simulation Experiments}
\subsubsection{Experiment Setup}
To evaluate the proposed general unbiased framework under different assumptions of user behaviors, we carry out a series of simulational experiments, i.e., simulating the users' click behaviors via specific probabilistic click models based on queries and documents data from public datasets.

\textbf{Dataset}. We use a publicly available LTR dataset MSLR-WEB30K \cite{QinL13}, which is generated by a commercial search engine and widely used in the evaluation of recommendation systems. The MSLR-WEB30K dataset contains more than 30000 queries with correspondingly preselected document lists (on average 125 documents per query). Each query-document pair is encoded by a feature vector, which takes the shape of (136,), and labeled by relevance, which takes five values from 0 (irrelevant) to 4 (perfectly relevant). In our setting, we binarize the relevance by only considering the document with a relevance value greater than two as relevant. 

\textbf{Click Simulation}. To simulate the production ranker in a real-world recommendation system, we follow existing works \cite{Oosterhuis20,Vardasbi20} and supervise train a non-optimal but decent production ranker using 1\% of the training data with relevance labels. 
Next, we use the gained production ranker to generate an initial document ranking list for each query, denoted as $\hat R_q$.
Then, we simulate the user browsing process and sample clicks for query $q$ based on the resulting document ranking $\hat R_q$ according to the observation probability and document relevance.
Specifically, the users' click behaviors are simulated under four assumptions:
\begin{itemize}
    \item[1.] users will click documents which are both observed and relevant;
    \item[2.] besides condition 1, users may click documents that are observed but irrelevant with a probability related to the item's presenting ranking;
    \item[3.] besides conditions 1 and 2, users may click documents that are relevant but not observed with a minor static probability (which is a common situation in applications like route recommendation and QAC);
    \item[4.] besides conditions 1, 2, and 3, users may click documents that are neither observed nor relevant with a minor static probability;
\end{itemize}
In addition, the observation probability are simulated under two assumptions:
\begin{itemize}
    \item[1.] the observation probability of an item depends on its presenting ranking;
    \item[2.] the observation probability of an item depends on its presenting ranking and a title-length feature (one of the features in query-document feature vector);
\end{itemize}
Note that each observation probability assumption can combine with any click behavior assumption, resulting in 8 composites. 
In our simulation experiments, we only simulate five composite scenarios S1-S5, in which S1-S4 all follow the first observation probability assumption and the first to fourth click behavior assumptions, respectively, while S5 follows the second observation probability assumption and the third click behavior assumption.

The following paragraphs will formulate the click simulation processes in the five scenarios mentioned above (S1-S5).

The click simulation in scenarios S1-S4 aims to evaluate our approach's adaptation towards different click generating conditions when the bias only depends on the position.
For scenario S1-S4, the probability of being observed, donated as $P(o_d=1|q,d,pos_d)$, is defined as follows:
\begin{equation}\label{eq:o1}
    P(o_d=1|q,d,pos_d)=\left\{
        \begin{array}{rcl}
            \frac{1}{pos_d},& &\text{if}\,\,pos_d\leq 5,\\
            0.1,& & \text{otherwise},
        \end{array}
    \right.
\end{equation}
Then the probability of a click, denoted as $P(c_d=1|q, d, pos_d , r_d, o_d)$ ($P(c_d=1|pos_d,r_d,o_d)$ for short), is conditioned on the relevance of the document according to the dataset besides observation probability, and defined based on four assumptions respectively as follows:
\begin{equation}\label{eq:clk1}\small
    P(c_d=1|pos_d,r_d,o_d)=
    \left\{
        \begin{array}{ll}
            1,&\text{if}\,\,r_d=1 \land o_d=1\\
            0,&\text{otherwise}
        \end{array}
    \right.
\end{equation}
\begin{equation}\label{eq:clk2}\small
    P(c_d=1|pos_d,r_d,o_d)=
    \left\{
        \begin{array}{ll}
            1,&\text{if}\,\,r_d=1 \land o_d=1 \\
            \frac{1}{\min(pos_d,\,5)+3},& \text{if}\,\, r_d=0 \land o_d=1\\
            0,&\text{otherwise}
        \end{array}
    \right.
\end{equation}
\begin{equation}\label{eq:clk3}\small
    P(c_d=1|pos_d,r_d,o_d)=
    \left\{
        \begin{array}{ll}
            1,&\text{if}\,\,r_d=1 \land o_d=1 \\
            \frac{1}{\min(pos_d,\,5)+3},& \text{if}\,\, r_d=0 \land o_d=1\\
            0.1,&\text{if}\,\, r_d=1 \land o_d=0\\
            0,&\text{otherwise}
        \end{array}
    \right.
\end{equation}
\begin{equation}\label{eq:clk4}\small
    P(c_d=1|pos_d,r_d,o_d)=
    \left\{
        \begin{array}{ll}
            1,&\text{if}\,\,r_d=1 \land o_d=1 \\
            \frac{1}{\min(pos_d,\,5)+3},& \text{if}\,\, r_d=0 \land o_d=1\\
            0.1,&\text{if}\,\, r_d=1 \land o_d=0\\
            0.01,&\text{otherwise}
        \end{array}
    \right.
\end{equation}

The click simulation in scenario S5 aims to evaluate our approach's adaptation towards different observation probability assumptions.
In scenario S5, the probability of being observed is defined considering the effect of the document itself besides the presenting ranking:
\begin{equation}\label{eq:o2}
    P(o_d=1|q,d,pos_d)=\left\{
        \begin{array}{lcl}
            \frac{\omega_\theta(d)}{pos_d},& &\text{if}\,\,pos_d \leq 5,\\
            0.1\omega_\theta(d),& & \text{otherwise},
        \end{array}
    \right.
\end{equation}
where $\omega_\theta$ is a linear function parameterized by $\theta$, and $\omega_\theta$ is normalized to [0.5, 1] for document list of each query using min-max normalization. 
Moreover, the probability of a click in S5 follows Eq.~\ref{eq:clk3}.

Using equations defined above, for each scenario, we simulate more than 10e6 clicks on all queries in the train set of MSLR-WEB30K/Fold1 dataset, and each query is sampled 66.5 times on average. We gather all document-click pairs during simulation to form the training set, while we directly use the document-relevance pairs in the valid set of MSLR-WEB30K/Fold1 dataset as testing set to evaluate all methods' ability on relevance inference.
Since 97\% of click labels in the training set are negative according to the above simulation progress, to balance the click label, we resample the training set before training according to a hyper-parameter $K$, which represents the additional number of documents sampled in each query's document ranking list besides the first five documents. In other words, at each click simulation iteration for a given query, we only gather the document-click pairs of the first five documents, and $K$ additional document-click pairs of $K$ randomly selected documents whose ranking is larger than 5. The value of $K$ is determined through an experiment of training a skyline model, in which a 3-layer MLP model is supervised trained by document feature vectors, which are in the $K$-sampled training set of assumption 1, and corresponding relevance labels. Moreover, we choose a rational $K$, which ensures the skyline model will not overfit after 70k training iterations. 
Eventually, $K$ is set to 15, and about 37 million document-click pairs are sampled for each scenario. 

Based on the simulation data under different assumptions, we first train different models. The batch size is set to 256, 
Then we evaluate and compare the AUC metric of different methods in the relevance prediction on the testing set.
We will introduce two compared state-of-the-art methods in the following paragraph.

\textbf{Compared Methods}.
\begin{itemize}
    \item MMoE-bias \cite{zhao2019recommending}: the debiasing framework used in this work models the positional bias by a shallow side tower independently and estimates the user click by adding logit for user engagement and logit for positional bias.
    \item PAL \cite{guo2019pal}: this approach models the probability of examination by a model, while the click rate conditioned on \emph{seen} by another model, and the final click rate is calculated by multiplying the outputs of two models.
\end{itemize}

\subsubsection{Results and Analysis}

\begin{figure*}[t]
    \centering
    \includegraphics[width=\textwidth,keepaspectratio]{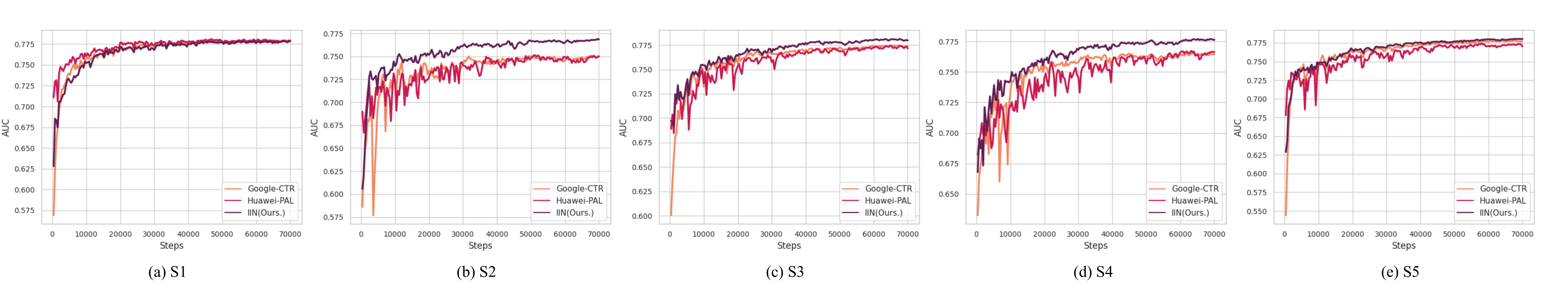}
    \caption{The changes of AUC on the test set with different methods during training progress.}
    \label{fig:auc}
\end{figure*}
This section presents and analyzes the comparison results between two baselines and our approach under scenarios S1-S4, where the bias depends only on position and user click behaviors following four different assumptions, respectively. Then we show and compare the results with different methods in scenario S5, where bias depends on the position factor and other features $b$ belonging to $x$.

In scenarios S1-S5, the change trends of AUC on the test set during the training process are shown in Figure \ref{fig:auc}. (a)-(e), respectively. 
Moreover, the AUC values achieved by three models after 70k training steps are listed in Table \ref{tab:auc}.

The results in scenarios S1-S4 show that, 
when the bias depends only on position, 
under the first condition, IIN achieves a similar AUC with PAL (-0.0011).
IIN outperforms MMoE-bias (+0.018, +0.0061, +0.0117) and PAL (+0.0184, +0.0082, +0.0102) under the last three click behavior generation conditions. 

Specifically, the performance of the three models is almost equivalent in scenario S1. The reason for the slight gap between IIN and PAL is that: 
in S1, clicks only generate when the item is relevant and observed, but IIN tries to estimate both click rates when the item is relevant or irrelevant, which is more complicated than estimating a single click rate value conditioned on seen like PAL.
Note that in S1, the click rate conditioned on seen estimated by PAL is equivalent to the probability of being relevant, and thus PAL achieves good AUC.

In contrast, in S2-S4, the click may generate under more conditions, in which PAL and MMoE-bias cannot effectively estimate the relevance between a query and a document.
For example, in S2, PAL predicts the bias, i.e., the probability of \emph{seen}, as well as the click rate conditioned on \emph{seen}. However, either relevant or irrelevant, a document may be clicked in S2 when seen, which means the estimated conditional click rate is not equivalent to the probability of being relevant anymore, leading to a drop in the AUC metric. In contrast, IIN can deal with this condition because it takes relevance as a latent variable and independently predicts the relevance probability.
Formally, the true value of click rate in S2 follows the form of $y=s^*(pos)r^*(x)+s^*(pos)(1-r^*(x))c^*(pos)$, in which $s^*$, $r^*$, and $c^*$ are all any functions.
For PAL, it estimates in a form of $y'=s(pos)c(x)$; for MMoE-bias, it estimates the true value in a form of $y'=b(pos)+r(x)$; while for IIN, it estimates the true value in a form of $y'=c_1(pos)r'(x)+c_2(pos)(1-r'(x))$, where $s,c,b,r,r',c_1,c_2$ are all any functions. It is easy to find that the function forms between true value and IIN prediction are equivalent, while the expression ability of PAL/MMoE-bias's true value estimation functions are insufficient.

We also note that in S3 and S4, PAL and MMoE-bias' performance has increased than performance in S2. The reason may be that a larger proportion of click data is generated on query-related documents in S3/S4 than S2.

The results in scenario S5 show that IIN also outperforms MMoE-bias (+0.0031) and PAL (+0.0067) when the bias depends on the document's position and title length, manifesting good generality of IIN.
 
\begin{table}[h]
\centering
    \begin{tabular}{|c|c|c|c|c|c|}
    \hline
    \textbf{AUC} & \multicolumn{1}{c|}{\textbf{S1}} & \multicolumn{1}{c|}{\textbf{S2}} & \multicolumn{1}{c|}{\textbf{S3}} & \multicolumn{1}{c|}{\textbf{S4}} &
    \textbf{S5} \\ \hline
    MMoE-bias  & 0.7790 & 0.75 & 0.7737 & 0.7646 & 0.7772 \\ \hline
    PAL  & \textbf{0.7794} & 0.7496 & 0.7716 & 0.7661  & 0.7736 \\ \hline
    IIN   & 0.7783 & \textbf{0.7684} & \textbf{0.7798} &  \textbf{0.7763} & \textbf{0.7803}\\ \hline
    \end{tabular}
    \caption{The AUC of different methods on CTR prediction under five scenarios:
    (1) S1-S4 is $pos$ determined bias, where S1 is observed$\&$relevant, S2 is S1+observed$\&$irrelevant, S3 is S2+unobserved$\&$relevant, and S4 is S3+unobserved$\&$irrelevant; (2) S5 is $pos+b$ determined bias, which is S3+$b$.}
    \label{tab:auc}
\end{table}


\subsection{Online Experiments}
\subsubsection{Online Experiments Setup}
Following an existing work\cite{cheng2021r4}, we design online experiments in a live recommendation system to verify the performance of IIN. 
Specifically, we conduct an eight days A/B test in a route recommendation scenario \cite{cheng2021r4} to validate the superiority of IIN over the current online baseline model, i.e., MMoE-bias.

\textbf{Route Recommendation System}.
In the route recommendation system, every time the user initiates a navigation request, he will get three recommended routes presented in an orderly manner.
10\% randomly selected users participated in the A/B test, and the number of navigation requests in the online A/B test is 500k per day.

\textbf{Offline Model Training}. 
Before the A/B test, the MMoE-bias baseline and IIN are trained with the same dataset generated from the online logs within 15 days.
In this dataset, for a candidate route in navigation, the input $x$ is the embedding of the route and user ($1 \times 186$); $b$ is the ($1 \times 26$) vector of features related to personal bias, such as user's historical yaw rate (a higher yaw rate indicates a higher bias); $pos$ is a ($1 \times 4$) one-hot vector; and the label $y$ is the degree of similarity between the candidate route and the actual driving route, named ACR. and defined by the following equation:
\begin{equation}
    y=\text{ACR}(\mathcal{P}, \mathcal{P}^*) = \frac{|\mathcal{P} \cap \mathcal{P}^*|}{|\mathcal{P}^*|}
\end{equation}
where $\mathcal{P}^*$ denotes the user's actual driving route during this navigation request, $\mathcal{P}$ denotes the candidate route, $\mathcal{P}_1 \cap \mathcal{P}_2$ denotes the sections where two routes intersect, and $|\mathcal{P}|$ denotes the length of a route $\mathcal{P}$.

The IIN model used in this offline training is an adapted version of the model shown in Fig.~\ref{fig:module}.(b). 
Specifically, we use a list-wise soft CE loss calculated on the list of recommended routes under a navigation request in online experiments instead of the point-wise CE loss calculated on each query-document pair used in simulation experiments.
The constraint used in the loss function is also replaced by the sum of each route's constraint value in the list.
For the MMoE-bias model, we also use two 3-layer MLPs with sigmoid regulations to estimate each route's bias score and relevance score, respectively. Moreover, the predicted output is calculated during training by adding a bias score and relevance score according to the MMoE-bias framework. The same list-wise soft CE loss function is used to train the MMoE-bias model without constraints.

\textbf{AB Test Settings}. For the control group, all users in the system are presented with route recommendations generated by the MMoE-bias framework. For the experimental group, all users in the system are presented with route recommendations generated by the IIN. 
Due to resource limits, The baseline and IIN are sequentially deployed in eight days instead of simultaneously: baseline online from July 28 to July 31, and IIN online from August 1 to August 4.

\textbf{Metrics} We adopt two metrics to evaluate and compare the online performance of IIN and baseline, namely the average \emph{first route coverage rate ($\overline{\text{FCR}}$)} and \emph{yaw rate ($\overline{\text{YR}}$)}.
\begin{definition}[Avg. First Route Coverage Rate] The average first route coverage rate, denoted as $\overline{\text{FCR}}$, depicts the average similarity between the first recommended route and the user's actual driving route in all navigation requests. In other words, this metric indicates the system's ability to rank the user's favorite route at the first position. $\overline{\text{FCR}}$ is defined as follows:
\begin{equation}
    \overline{\text{FCR}} = \frac{1}{N}\sum_{v=1}^N \text{FCR}(\hat{R_v},\mathcal{P}_v^*)=\frac{1}{N}\sum_{v=1}^N ACR(\mathcal{P}_v^1, \mathcal{P}_v^*)
\end{equation}
where $N$ indicates the total number of navigation, $\hat{R_v}$ denotes the recommended list of routes in the $v^{th}$ navigation, $\mathcal{P}_v^l \in \hat{R_v}$ denotes the $l^{th}$ recommended route in the list, and $\mathcal{P}_v^*$ denotes the user's actual driving route during the $v^{th}$ navigation.
\end{definition}
\begin{definition}[Avg. Yaw Rate] The average yaw rate, denoted as $\overline{\text{YR}}$, depicts the probability that the user deviates from all recommended routes during driving, therefore indicating an overall effect of route recommendation. $\overline{\text{YR}}$ is defined as follows:
\begin{equation}
    \overline{\text{YR}} = \frac{1}{N} \sum_{v=1}^N \mathbf{1}(|\mathcal{P}_v^* \setminus \bigcup_{l=1}^{|\hat{R_v}|} \mathcal{P}_v^l|>0)
\end{equation}
where $|\hat{R_v}|$ indicates the number of routes in the recommendation list, and $\mathbf{1}(x) = 1$ when $x$ is True, otherwise 0. Therefore, $\bigcup_{l=1}^{|\hat{R_v}|} \mathcal{P}_v^l$ denotes the road network formed by all recommended routes, and when this network cannot cover user's actual driving route, the length of the differentiate set $\mathcal{P}_v^* \setminus \bigcup_{l=1}^{|\hat{R_v}|} \mathcal{P}_v^l$ is greater than 0.
\end{definition}

\subsubsection{Results and Analysis}

In the A/B test, both $\overline{\text{FCR}}$ and $\overline{\text{YR}}$ are evaluated under two categories of navigation: short-distance navigation ($\leq$2km) and long-distance navigation (>2km). Furthermore, under both categories, users are divided into three groups: 
\begin{itemize}
    \item[u1.] users who are unfamiliar with the navigation route and have not set route preferences.
    \item[u2.] users who are unfamiliar with the navigation route and have set route preferences.
    \item[u3.] users who are familiar with the navigation route.
\end{itemize}
The $\overline{\text{FCR}}$ and $\overline{\text{YR}}$ results in A/B test for different navigation scenarios abd user groups are shown in Table~\ref{tab:metrics}.



\begin{table}[h]
\centering
    \begin{tabular}{|c|c|c|c|c|c|c|}
    \hline
    \multirow{2}{*}{Metrics} & \multicolumn{3}{c|}{Short Dis.} & \multicolumn{3}{c|}{Long Dis.} \\\cline{2-7}
    & u1 & u2 & u3 & u1 & u2 & u3 \\
    \hline
    $\overline{\text{FCR}}$(\%)  & \textbf{+1.46} & +1.21 & +0.49 & \textbf{+1.82}  & +0.75 & +0.18\\ \hline
    $\overline{\text{YR}}$(\%)  & \textbf{-1.53} & -0.88 & -0.30 & \textbf{-1.32} & -0.83 & -0.50 \\ 
    \hline
    \hline
    $\overline{\text{FSR}}$(\%)  & \textbf{98.94} & 98.88 & 97.78 & \textbf{95.76} & 95.75 & 93.10 \\
    \hline
    \end{tabular}
    \caption{The improvements on $\overline{\text{FCR}}$ and $\overline{\text{YR}}$ metrics under different cases in online A/B test, as well as the value of $\overline{\text{FSR}}$ under different cases.}
    \label{tab:metrics}
\end{table}

These results show that:
(1) under all user groups in both short distance and long distance navigation, IIN outperforms MMoE-bias baseline in both $\overline{\text{FCR}}$ and $\overline{\text{YR}}$ metrics;
(2) for $\overline{\text{FCR}}$, IIN achieves the largest $\overline{\text{FCR}}$ improvement with the user group u1 in both short and long distance navigation (+1.46\% and +1.82\%);
(3) for $\overline{\text{YR}}$, IIN achieves the largest $\overline{\text{FCR}}$ reduction with the user group u1 in both short and long distance navigation (-1.53\% and -1.32\%).

The last line in Table.~\ref{tab:metrics} also shows the average probability of a kind of user behavior, i.e., clicking the first recommended route directly, namely \emph{first route selection rate} ($\overline{\text{FSR}}$), within each user group in both short and long-distance navigation. These statistical results imply that users within an unfamiliar scenario (u1) have higher positional bias, requiring an effective debiasing method.
Besides, although it is not convenient for us to disclose all the absolute values of metrics due to commercial privacy, it is worth noting that in both short and long-distance navigation, the metrics in u2/u3 are obviously greater than u1.
Because users in groups u2 and u3 usually provide more explicit bias-related features to the recommendation system, like the historical yaw rate, while in user group u1, the bias-related features are insufficient.
Surprisingly, facing higher debiasing needs and more limited bias-related features, IIN achieves the maximal improvements with the user group u1 in both short and long-distance navigation, manifesting the effectiveness of our method in debiasing. 

\begin{figure}[t]
    \centering
    \includegraphics[width=0.45\textwidth,keepaspectratio]{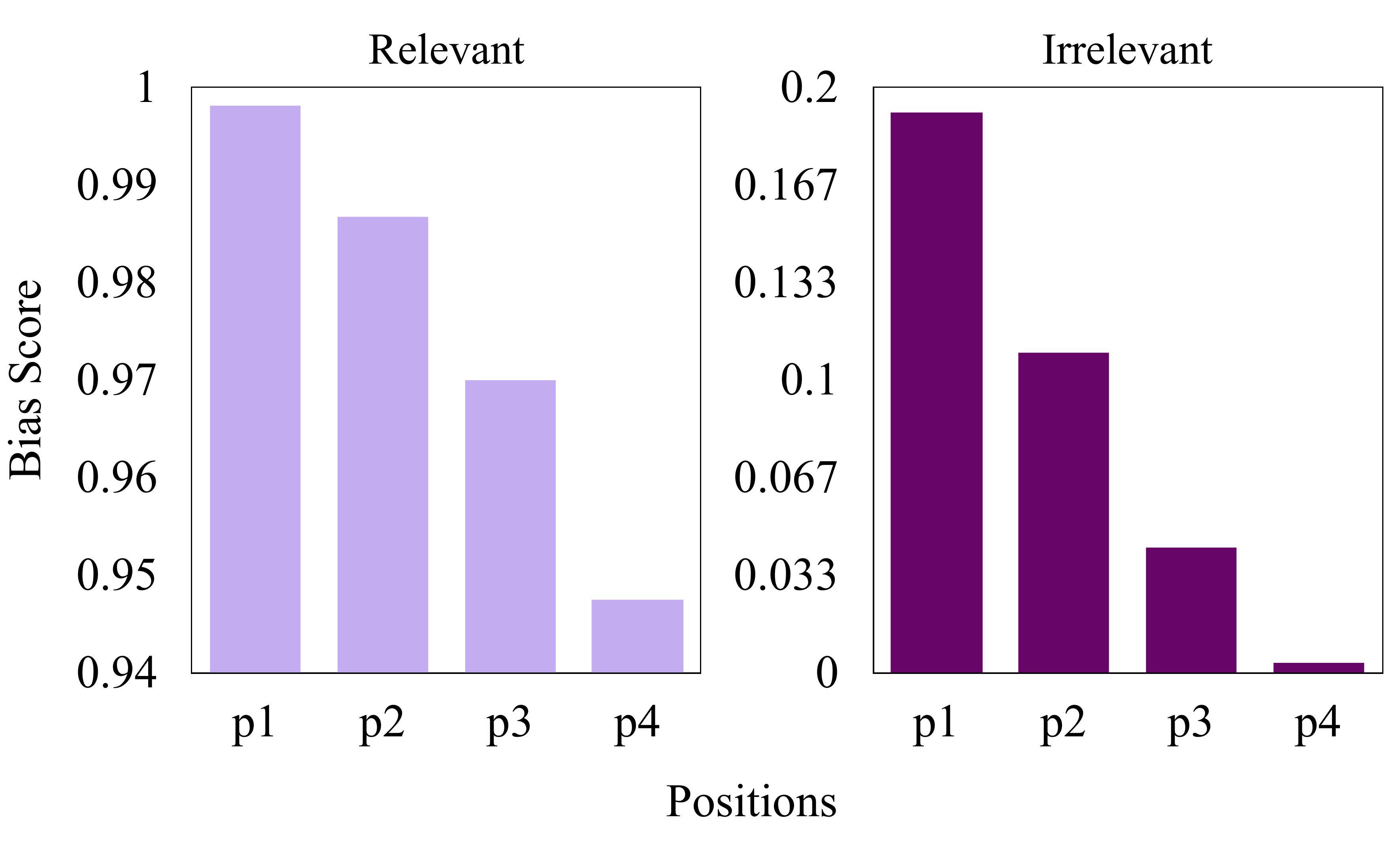}
    \caption{The estimated bias score at different positions when the route is either relevant or irrelevant.}
    \label{fig:bias}
\end{figure}

To further verify the debiasing ability of our approach, we statistic and visualize the estimated bias score in the online experiments.
Specifically, in online experiments, the bias on each of the 4 position outputs by the IIN bias module is a $2\times2$ matrix $\begin{bmatrix}
p(c=1|r=1) & p(c=1|r=0) \\
p(c=0|r=1) & p(c=0|r=0)
\end{bmatrix}$. Here, we only visualize the value of $p(c=1|r=1)$ and $p(c=1|r=0)$ on 4 positions, respectively. 
Results are shown in Figure~\ref{fig:bias}.
As expected, no matter the route is relevant or irrelevant, the position with a higher ranking has a higher bias score, which leads to effective reweight and debiasing of route score estimation.

\section{Conclusion}
Debasing is a critical problem in click-through rate prediction, but existing methods depend on an oversimplified assumption and are insufficient to model the rich interaction between position and other information. 
In this paper, we argued that given different users and items, there are four scenarios: 1. observed $\&$ relevant, 2. ob-served $\&$ irrelevant, 3. unobserved $\&$ relevant, and 4. unobserved $\&$ irrelevant.
Therefore, we proposed a general framework, named IIN, for debiasing in CTR prediction to deal with all four scenarios.
Specifically, based on a perspective of a probabilistic graphical model, we designed two neural networks in the framework to estimate the user's interest and the transition probability of clicking, respectively. Online inference only used the interests predicting module.
We also carried out both offline and online experiments.
The experiment results demonstrated the superiority of IIN compared with two state-of-the-art counterparts.

\bibliographystyle{coling}
\bibliography{coling2020}

\end{document}